# Search for Active and Inactive Ion Insertion Sites in Organic Crystalline Materials

Harshan Reddy Gopidi,[1,2,3] Alae Eddine Lakraychi,[1,2,3] Abhishek A. Panchal,[1,2,3] Yiming Chen,[1,4] V. S. Chaitanya Kolluru,[1,4] Jiaqi Wang,[1,5] Ying Chen,[1,6] Liu Jue,[7] Kamila Wiaderek,[1,5] Maria K.Y. Chan,[1,4] Yan Yao,[1,2,3,*] and Pieremanuele Canepa[1,2,3,#]

[1]Energy Storage Research Alliance, Argonne National Laboratory, 9700 South Cass Avenue, Lemont, IL 60439, USA
[2]Texas Center for Superconductivity, University of Houston, Houston, TX 77204, USA
[3]Department of Electrical and Computer Engineering, University of Houston, Houston, TX 77204, USA
[4]Center for Nanoscale Materials, Argonne National Laboratory, Lemont, IL 60439, USA
[5]X-ray Science Division, Argonne National Laboratory, Lemont, IL 60439, USA
[6]Physical and Computational Sciences Directorate, Pacific Northwest National Laboratory, Richland, WA 99354, USA
[7]Neutron Sciences Directorate, Oak Ridge National Laboratory, Oak Ridge, TN 37831, USA

*email: yyao4@central.uh.edu
#Email: pcanepa@uh.edu

# Abstract

The position of mobile active and inactive ions, specifically ion insertion sites, within organic crystals, significantly affects the properties of organic materials used for energy storage and ionic transport. Identifying the positions of these atom (and ion) sites in an organic crystal is difficult, especially when the element has a low X-ray scattering power, such as lithium (Li) and hydrogen, which are difficult to detect with powder X-ray diffraction (XRD) methods. First-principles calculations, exemplified by density functional theory (DFT), are very effective for confirming the relative stability of ion positions in materials. However, the lack of effective strategies to identify ion sites in these organic crystalline frameworks renders this task extremely challenging. This work presents two algorithms: (i) Efficient Location of Ion Insertion Sites from Extrema in electrostatic local potential and charge density (ELIISE), and (ii) ElectRostatic InsertioN (ERIN), which leverage charge density and electrostatic potential fields accessed from first-principles calculations, combined with the Simultaneous Ion Insertion and Evaluation (SIIE) workflow –that inserts all ions simultaneously—to determine ion positions in organic crystals. We demonstrate that these methods accurately reproduce known ion positions in 16 organic materials and also identify previously overlooked low-energy sites in tetralithium 2,6-naphthalenedicarboxylate ($Li_4NDC$), an organic electrode material, highlighting the importance of inserting all ions simultaneously as done in the SIIE workflow.

## Introduction

Understanding the structure of inorganic and organic-based materials containing electroactive elements, such as Li, Na, and zinc (Zn)[1–14] is relevant for the development of the next generation of rechargeable battery technologies, which are prevalent in vehicular transportation, heavy-duty applications, and the supporting electrical grids.

With the advent of the Rietveld methods,[15–18] diffraction-based techniques using X-ray (XRD) or neutron (ND) sources are commonly employed to determine the structures of energy materials. However, structural determinations through diffraction methods can be hindered by the weak scattering of X-rays by light elements, such as hydrogen and lithium (Li), which can partially occupy multiple crystallographic sites because of their high intrinsic mobility. In some cases, significant ion mobilities can cause the blurring of Bragg intensities.[19,20] While ND experiments enhance sensitivity to specific elements, e.g., light elements H and Li, ND experiments are significantly less available than XRD, and they still face challenges related to disorder, defects, and temperature-dependent site mixing.[21]

Organic electrode materials (OEMs) are promising for inexpensive and high-energy-density rechargeable batteries. Indeed, OEMs can provide three main benefits: (i) high material-level energy densities of 900-1000 kWh kg$^{-1}$, (ii) a growing range of designs as shown by the variety of molecules studied so far, and (iii) potential benefits in sustainability and supply chains.[6,22–26] One of the main challenges hindering the practical implementation of OEMs is a poor understanding of their structure-property relationships. The electrochemical reaction pathways of OEMs are typically interpreted based on simplistic molecular models, rather than describing these processes at the material level. These molecular-type OEM models are often paired with *ex situ* or *operando* XRD experiments, which frequently cannot reveal their underlying crystal structures and potential phase transformations.[27–30] To date, only a handful of studies have resolved structures at both charged and discharged (the ion-containing form) states of specific OEMs.[23,31–35] For example, 1,4-benzoquinone (BQ) undergoes a phase transition from monoclinic (*P2$_1$/c*) to orthorhombic (*P4$_2$/ncm*) upon lithiation or sodiation, forming Li$_2$BQ or Na$_2$BQ,[23,36] indicative of conversion-type reactions. By contrast, dilithium 2,6-naphthalene dicarboxylate (Li$_2$NDC) retains a monoclinic (*P2$_1$/c*) structure upon complete lithiation (Li$_4$NDC),[32,33,37] following an intercalation reaction. However, distinguishing between specific reaction mechanisms remains poorly understood in the OEM literature, mainly because of the lack of structural data for the discharged state, unlike the charged state, which is often known.

A comprehensive understanding of OEM structures and their behavior across different ion content (i.e., different states of charge) remains crucial for bridging the gap between molecular-scale insights and crystalline materials with intrinsic periodicity. In general, this knowledge is essential for elucidating the physicochemical processes that govern electrochemical behavior, including phase transitions,[23] voltage profiles,[35] and mechanical degradation.[34,38] However, accurate identification of the positions of mobile ions such as Li$^+$, Na$^+$, K$^+$, Zn$^{2+}$, and Mg$^{2+}$, in organic crystalline frameworks, remains a significant experimental and computational challenge.[27,31,33,34,39–47] This challenge arises primarily from the ample intermolecular space (void within the crystal framework) caused by weak intermolecular interactions, which are responsible for molecular assembly into organic molecular crystals. This results in numerous possible cation arrangements within the organic framework. Therefore, developing computational tools that can accurately predict ion positions within the crystalline structure of organic materials is essential for making breakthroughs and unlocking hidden potential for rational materials design from a crystal-level perspective. While such approaches have been established for inorganic electrode materials, [48–51] predictive search for ion sites in OEMs remains largely confined to the molecular scale and,[44,52] to the best of our knowledge, has not yet been demonstrated for crystalline organic frameworks.

In this paper, we present two complementary computational algorithms and an ion insertion workflow for identifying the optimal active and inactive ion insertion sites within organic frameworks: (i) The Efficient Location of Ion Insertion Sites from Extrema (ELIISE) in electrostatic local potential and charge density, which identifies electrostatically stable ion positions by locating local extrema in electrostatic local potential and electronic charge density (CD) fields. (ii) The ElectRostatic InsertioN (ERIN), which, by following the electrostatic potential, iteratively places ions into electrostatically favorable regions of the unit cell until the entire cell volume is sampled. (iii) The Simultaneous Ion Insertion and Evaluation (SIIE), which inserts the ions simultaneously and evaluates to determine the correct structure. ELIISE and ERIN, in conjunction with SIIE, are applied with a library of 16 representative organic structures, illustrating their ability to automatically identify candidate ion sites within complex, flexible organic frameworks using limited computational resources. These predictions are subsequently refined using agnostic first-principles calculations, resulting in high-confidence ion positions, which provide crucial insights into ion coordination environments and other structural evidence. These predictive workflows establish a powerful toolkit for the systematic exploration of electroactive structures and their ion transport pathways, elucidating reaction mechanisms and guiding the rational design of materials for the development of competitive organic-based electrochemical devices, as well as the exploration of soft materials for other energy-storage applications.

## Methods

### 1. Finding candidate ion insertion sites in organic crystals: ELIISE and ERIN

**Figure 1** shows the identification of candidate ion sites using ELIISE and ERIN, which rely on the electronic charge density (CD, or $\rho(r)$) and the local electrostatic potential $LP(r)$ **Eq. 1**, and the spherically averaged local potential, $\mathrm{ALP}(r)$, of **Eq. 2**.

$$LP(r) = V_{hartree}(r) + V_{ionic}(r) \qquad \text{Eq. 1}$$

where $V_{hartree}(r)$ is the Coulomb potential set by $\rho(r)$, and $V_{ionic}(r)$ is the Coulomb potential originating from the ionic core of all atoms. From $LP(r)$, the $ALP(r)$ is defined as:

$$\mathrm{ALP}(r) = \frac{\int_{V_s} LP(r+r')\,d^3 r'}{\frac{4}{3}\pi r_s^3} \qquad \text{Eq. 2}$$

where $V_s$ represents the spherical region within radius $r_s$ that is averaged at each point $r$, and $r'$ is a dummy variable over which the averaging occurs. Note, in **Eq. 2**, $ALP(r)$ is the spherically averaged electrostatic potential at each point in the organic crystal framework, with constant radii ($r_s$) set by the user (here, fixed to the empirical covalent radii of the atom that is inserted).

Both methodologies rely on an initial organic framework to compute $LP(r)$ and CD. The choice of the initial organic framework is trivial when the organic framework of the discharged (ion-inserted) structure— i.e., the lattice parameters, the symmetry, and the atomic positions— is known. In such cases, only the precise determination of ion positions is necessary, as is often the case with lithium; here, ELIISE serves as an appropriate tool. Conversely, the more prevalent scenario is when only the organic framework of the charged (ion-removed) phase is known. Under these circumstances, ERIN is recommended, as employing ELLISE may result in unreliable outcomes.

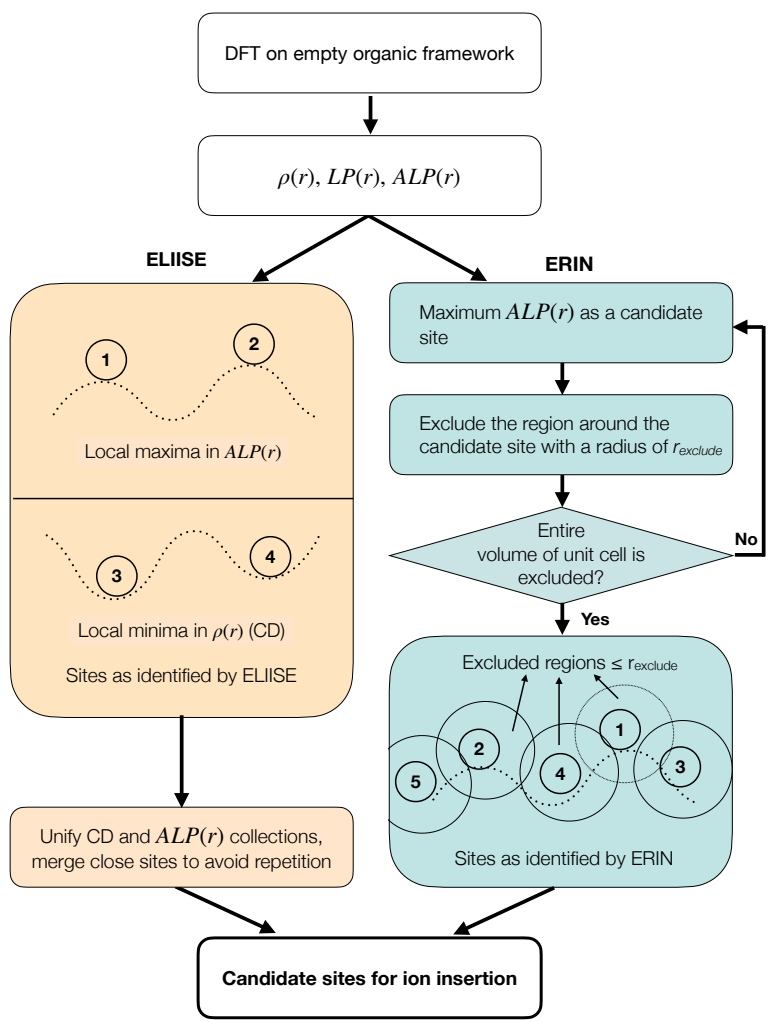

**Figure 1: Workflow and schematic illustration of the ELIISE (left) and ERIN (right) algorithms.**
ELIISE: Process for identifying potential ion sites based on local potential and charge density. ERIN: Workflow illustrating the iterative process used to identify potential ion sites based on electrostatic stability systematically.

## 2. The Efficient Location of Ion Insertion Sites from Extrema in electrostatic local potential and charge density (ELIISE) Method

The ELIISE method identifies candidate ion sites using both the CD and the spherically averaged local potential $ALP(r)$, defined in **Eq. 2**, within the empty (i.e., without any mobile species) organic framework from the discharged state.

Following the ELIISE workflow in the left part of **Figure 1**, the local potential and electronic CD are initially computed by performing a first-principles calculation without changing the atom positions (or changing the volume or cell shape of the material) of the empty organic framework. Subsequently, ELIISE identifies candidate active sites in an organic framework by identifying the local maxima and the local minima in the $ALP(r)$ the CD fields, respectively. **Figure 1** shows a schematic diagram of the local minima in CD and local maxima in the $ALP(r)$, which are used to identify candidate sites. We have concluded that a spherical average of the $LP(r)$ provides the best predictions in terms of candidate ion sites, as it accounts for the finite size of inserted ions. Here, the empirical covalent radii of atoms by Slater[53] appear to be a reliable option for defining this ion size. When different types of active species are involved in ion insertion, we imposed the empirical radius of the smaller atom for spherical averaging of the local potential. The candidate sites identified by both the CD and the $ALP(r)$ descriptors are combined into a single collection. To avoid repetition of sites, spatially close sites are merged by calculating the geometric mean of their coordinates, as indicated in **Figure 1**. The resulting sites are then ordered by their electrostatic stability, from the most favorable to the least favorable, as determined by the $ALP(r)$.

Large values of $ALP(r)$ indicate regions that are electrostatically favorable for positively charged ions, typically in the vicinity of more electronegative (negatively charged) atoms within the crystal structure. Therefore, local maxima in $ALP(r)$ naturally point to potential cation sites. Similarly, local minima in the CD usually occur in voids near more electronegative atoms. It is reasonable to assume that these electronegative atoms strongly attract electron density, leading to electron-deficient regions nearby, which are likely to form local minima in the charge density. Therefore, these minima of CD often align with favorable coordination environments for cations. An exception to this trend becomes clear when local minima occur in large structural voids, which are common in organic crystals or nanoporous materials. These voids may be too far from the organic framework to provide any stabilizing interactions for inserted ions. These situations can lead to surprisingly high Coulomb energies. To address this potential issue during the site search in both ELIISE and ERIN, we set cutoff distance thresholds to exclude regions that are either closer than the specified distance (e.g., 1.5 Å) or farther than ~3 Å from atoms in the organic framework.

### 3. The ElectRostatic InsertioN (ERIN) Method

In contrast to ELIISE, ERIN only requires the $LP(r)$ (**Eq. 1**) of the organic framework, obtained from first-principles calculations. The ERIN approach in the right part of **Figure 1** only uses the spherically averaged local potential $\text{ALP}(r)$ (**Eq. 2**) to explore electrostatic energy landscapes and systematically generates sparsely spaced candidate ion sites in a decreasing order of electrostatic favorability within the given unit cell. ERIN identifies and selects the site with the maximum $ALP(r)$ value. Subsequently, ERIN excludes a spherical region within a user-defined radius, r$_{\text{exclude}}$ (e.g., 2 Å), around this site from future searches. Then, ERIN repeats the search for the site with the next highest value of $ALP(r)$ to select as a candidate. This process is repeated iteratively until the entire volume of the unit cell is scanned systematically and excluded. The site search is performed independently within each symmetry-distinct region of the unit cell to identify candidate sites in all possible Wyckoff positions.

4. **Simultaneous Ion Insertion and Evaluation (SIIE) Workflow for Candidate Site Insertion**

Once ELIISE or ERIN determines the candidate sites, the next step is to identify the optimal combination of sites yielding the expected stoichiometry of the system and energetically viable structures. For this purpose, we introduce the Simultaneous Ion Insertion and Evaluation (SIIE) workflow, in which ions are introduced simultaneously within the organic framework to achieve the desired stoichiometry, as opposed to the sequential insertion method, where ions are inserted one at a time.[48,54] In the SIIE workflow, if the organic framework of the discharged structure is known, its atoms are kept fixed during ion insertion, with only inserted ions and decorating hydrogen atoms (whose positions are typically unknown from standard XRD experiments) allowed to change. Otherwise, if only the organic framework of the charged phase is known, then both the ions and the organic framework are allowed to relax completely.

As shown in **Figure 2a**, SIIE uses the candidate ion insertion sites from the ELIISE or ERIN methods discussed in the previous sections as inputs. While ELIISE and ERIN provide sets of candidate ion positions, they do not determine the final configuration of ions in the discharged structure. In real systems, multiple symmetry-distinct sites may be occupied simultaneously, and therefore, the correct discharged state often results from a specific combination of these candidate sites.

Finding the right combination of candidate sites is challenging due to the large number of unique possibilities. Note that the number of possible combinations of predicted ion sites that produce the target ion content can grow combinatorially, in the range $10 - 10^6$, which typically makes direct DFT evaluation of all combinations computationally infeasible.

To address this challenging task, we incorporate a machine learning interatomic potential (MLIP), such as MACE or Orb-v3,[55–57] to screen thousands of configurations efficiently. Here, we utilize MACE, which enables us to identify and remove high-energy structures. More details on how MACE and DFT are combined in the SIIE workflow are provided later (see **Computational Details**).

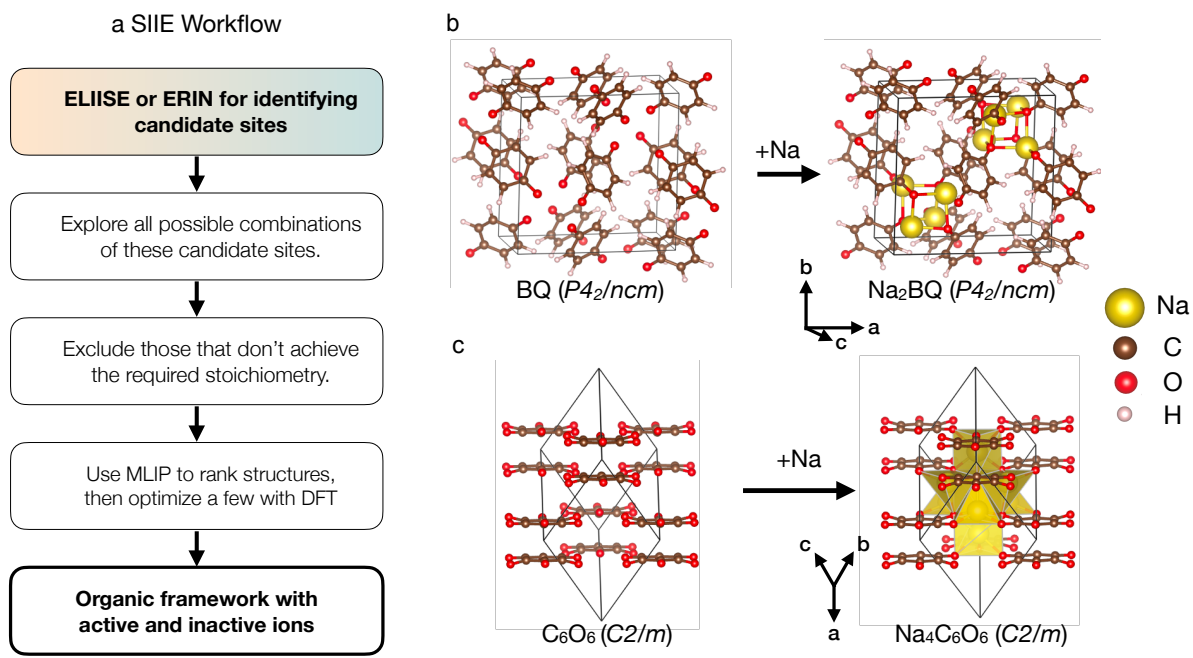

**Figure 2: Workflow for the Simultaneous Ion Insertion and Evaluation (SIIE) method (a) and its application in two organic-based materials.** (b) Disodium 1,4-benzoquinone ($Na_2BQ$) and (c) Tetrasodium rhodizonate ($Na_4C_6O_6$). Combining SIIE with ELIISE, the Na-ion positions are located in two exemplary electro-active organic materials.[36,58]

## Results

**Case Studies: Implementing the Suggested Workflows on Example Systems**

To demonstrate the effectiveness of these algorithms, we showcase the application of the ELIISE (ERIN) + SIIE workflow to several model compounds. The accuracy of ELIISE+SIIE is verified by "rediscovering" the Na-ion sites in two known fully sodiated compounds: (i) the disodium 1,4-benzoquinone (*P4$_2$/ncm*, Na$_2$BQ),[36] and (ii) the tetra-sodium rhodizonate with formula Na$_4$C$_6$O$_6$ (*C2/m*).[58] Similarly, starting from the known dilithium 2,6-naphthalenedicarboxylate (*P2$_1$/c*, Li$_2$NDC), using ERIN+SIIE, we predict Li positions in Li$_4$NDC (tetra-lithium 2,6-naphthalenedicarboxylate),[33,59] which contains 2 additional Li sites per formula unit (f.u.).

Na$_4$C$_6$O$_6$ (*C2/m*)[58] is an oxocarbon-salt-based layered compound; it is a planar conjugated ring with six carbonyl oxygens, known to undergo a two-to-four electron reduction with Na. To test the ELIISE method, we first remove all Na atoms from Na$_4$C$_6$O$_6$ and perform a static DFT calculation, at the experimentally determined volume and geometry, to calculate the electronic charge density and local electrostatic potential fields and subsequently apply the ELIISE algorithm to find the possible sites for Na-ions (as described in **Methods Sec. 2**).

The ELIISE algorithm yields nine symmetry-distinct Na sites with different multiplicities (**Table S2**). These candidate sites from ELIISE are then "funneled" into the SIIE workflow. We examine all possible combinations of sites from these nine symmetry-distinct candidates that produce the desired Na$_4$C$_6$O$_6$ stoichiometry. This yields only 59 unique sodium configurations, for which we perform DFT and MACE optimizations to identify the most plausible Na-atom positions. To quantify the prediction accuracy of ELIIER+SIIE, we use a metric called the maximum atom displacement error (MADE), and defined in **Eq. 3**:

$$\text{MADE} = Max\{|r_i - r_i^{ref}|;\ i \in \text{all active and inactive species}\} \quad \textbf{Eq. 3}$$

Where $r_i$ is the predicted ion position of the $i^{th}$ ion, and $r_i^{ref}$ is the position of the corresponding ion from the literature. The predicted and literature ion pairs are matched based on the pairing that minimizes the sum of their absolute deviations. Thus, MADE (**Eq. 3**) is the maximum displacement between any predicted ion position and its corresponding literature reference, i.e., the actual atom positions reported in the literature.[58] MADE is defined when the organic framework of the discharged structure is known and remains fixed during the optimization of ionic positions. MADE measures the accuracy of the predicted sites by indicating how much the ions deviate from literature positions.

In practice, we interpret MADE values ≤~0.4 Å as indicating correct predictions for all ions in the structure, while MADE values ≥~0.4 Å suggest that at least one ion site has been predicted incorrectly. Among the predicted structures for $Na_4C_6O_6$, the lowest-energy structure of **Figure 2c** displays a MADE of ~0.026 Å, indicating excellent agreement with the structure previously reported in the literature.[58]

In the second example, we investigate the structure of disodium 1,4-benzoquinone, $Na_2BQ^{36}$ (**Figure 2b**), and the fully reduced (sodiated) form of a para-benzoquinone, i.e., BQ. In Na$_2$BQ, the Na atom positions are available from the literature.[36] In **Figure 2b**, the Na$_2$BQ features a highly open organic framework with large voids, making a brute-force search of the potential Na positions computationally expensive. Using ELIISE, we identify four symmetrically distinct candidate sites. Applying the SIIE workflow, we evaluate all combinations of these four symmetry-distinct ions that produce the stoichiometry of Na$_2$BQ. This step yields only three valid configurations, which we used for further DFT structural relaxations (keeping the volume and shape fixed to the experimental structure of Na$_2$BQ). The most stable structure with the lowest DFT total energy singled out with ELIISE+SIIE, shown in **Figure 2b**, exhibits a MADE of ~0.024 Å, demonstrating excellent agreement with the experimental Na positions.[36]

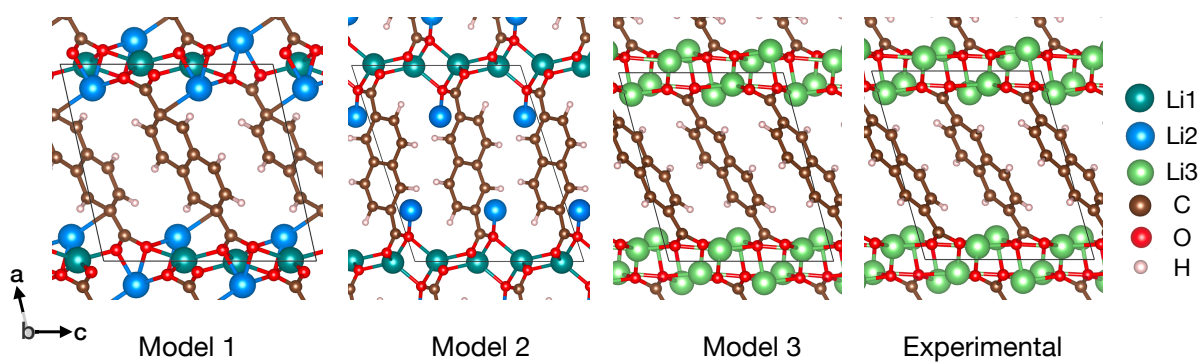

**Figure 3**: **Li$_4$NDC structures:** Model 1, Model 2, Model 3, and experimentally refined from single-crystal XRD.[33] Different crystallographic Li sites are shown.

We tested our methods on a third structure, Li$_4$NDC, which shares structural similarities to its precursor Li$_2$NDC,[60] with 2 Li atoms per molecule. Prior experimental studies indicated that the NDC organic framework largely remains unchanged upon reaction with Li. Two structural Li arrangements, Model 1 and Model 2 (**Figure 3**), have been proposed for Li$_4$NDC.[59] Models 1 and 2 preserve Li-ions in tetrahedral coordination sites as observed in Li$_2$NDC[60] (shown as Li1 in **Figure 3**) and introduce new Li positions (shown as Li2 in **Figure 3**), approaching composition Li$_4$NDC (four Li atoms per molecule).

Using ERIN with SIIE workflow, we predict the Li positions in Li$_4$NDC based on the organic framework of Li$_2$NDC as input. We recovered Models 1 and 2, described in the literature. A close inspection of Model 1 and Model 2 reveals that these two structures reported in Ref. [59] are different in the relative orientation of NDC molecules, as displayed in **Figure 3**. However, this striking contrast is not highlighted in Ref. [59], which seems to suggest that Model 1 and Model 2 differ only by Li distribution but maintain the same molecular framework.

Along with Models 1 and 2, we identified a different model as the lowest-energy structure, as determined by DFT minimization (and MACE), which we named Model 3. In Model 3, the tetrahedral sites reported for Li$_2$NDC are unoccupied, and all the Li atoms assume new crystallographic positions (collectively indicated as Li3 in **Figure 3**), which are localized near the inorganic layer of the organic framework, consisting of oxygen atoms in NDC. These newly identified Li-atom positions of Model 3 provide an electrostatically favorable arrangement of the Li atoms compared to Models 1 and 2, at the same composition. Quantitatively, DFT total energy comparisons show that Model 3 is energetically favored over both Model 1 and Model 2 by ~100 meV/Li$_4$H$_6$C$_{12}$O$_4$. and ~440 meV/Li$_4$H$_6$C$_{12}$O$_4$, respectively.

Indeed, Model 3 is in excellent agreement with a recent study (appeared while drafting this paper) that used single-crystal XRD to refine the structure of Li$_4$NDC. [33] Because ERIN uses the organic framework of the charged state as input and optimizes all atomic positions, the MADE metric—measuring the prediction error in active species positions—is not suitable. Instead, a simple qualitative measure of similarity between two structures can be obtained using, for example, StructureMatcher (using default parameters) as implemented in Pymatgen.[61] Here, StructureMatcher indicated that model 3 qualitatively matches the recently reported experimental structure of $Li_4NDC$.[33]

These three case studies, discussed in the previous paragraphs, collectively demonstrate the robustness of the ELIISE and ERIN methods when integrated with the SIIE workflow in accurately identifying ion positions across various organic electrode materials.

**Assessing the accuracy of the ELIISE+SIIE approach**

To assess the accuracy of the ELIISE + SIIE approach, we apply it to a diverse set of organic electrode materials (16 in total) for which the organic framework and ion positions of the crystal structure are known from literature. Using MADE (**Eq. 3**) as a metric for similarity, we evaluate the performance of the ELIISE method with the SIIE workflow across 16 organic systems. These structure organic electrode materials, such as $Li_2NDC$ and $Li_4NDC$ (di and tetra lithium 2,6-naphthalene dicarboxylate),[60] $Li_2BDC$ and $Li_4BDC$ (di and tetra lithium 1,4-benzene dicarboxylate),[62] $Li_2$-BPDC (dilithium biphenyl dicarboxylate),[63] $Li_2Mg$-$p$-DHT (magnesium(2,5-dilithium-oxy)-terephthalate),[31] $Na_2BQ$ (disodium hydroquinone),[36] $Na_2C_6O_6$ (disodium rhodizonate),[64] and $Na_4C_6O_6$ (tetra sodium rhodizonate),[58] as well as select structures (see **Table S1**) obtained from crystallographic databases, the Cambridge Crystallographic Data Centre (CCDC)[65] and the Inorganic Crystal Structure Database (ICSD).[66]

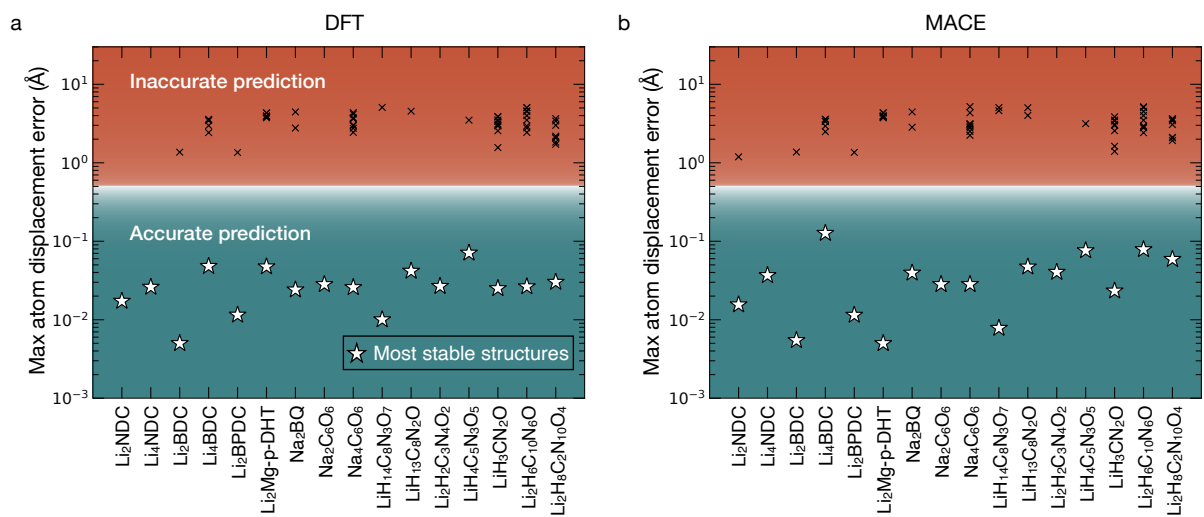

**Figure 4: Accuracy of the predicted active ion positions with the proposed SIIE workflow and ELIISE.** MADE (Eq. 3) values determined after optimization of ionic positions with (a) first-principles DFT and (b) the machine learned potential MACE. Stars indicate the most stable (lowest total energy) ion arrangements in each organic structure, and crosses indicate the thermodynamically unfavorable (high energy) structures. Details of reference structures reported in the literature are in **Table S1**.

As shown in **Figure 4a**, the lowest-energy structure identified through the SIIE workflow using ELIISE correctly reproduces the known ion positions in all tested cases, as measured by MADE, whose value is always ≤ 0.4 Å in all the cases considered. Furthermore, **Figure 4b** demonstrates that using a pre-trained machine learning interatomic potential (MLIP), specifically MACE (only used for optimizing ion positions), yields the correct prediction of ion positions, with MADE values consistently lower than 0.4 Å. This suggests that DFT-level structural relaxation may not be strictly necessary for determining the general location of inserted ions, leading to a substantial reduction in computational cost. **Figure S1** presents a similar assessment, but with predictions generated using ERIN+SIIE instead of ELIISE+SIIE, and only MACE was employed for optimizing ion positions. The MADE values in **Figure S1** show that ERIN+SIIE also accurately predicted the positions of all inserted ion sites for all 16 tested organic systems.

## Discussion

Predicting ion positions in "soft" crystalline materials, such as organic electrode materials, remains a significant challenge due to the combination of large free volumes, weak intermolecular forces, and the dynamic nature of ion coordination. In this paper, we have developed two new methodologies, ELIISE and ERIN, along with an ion insertion workflow, SIIE, designed to identify ion insertion sites in organic structures. These include metal atoms, such as Na, Zn, and Mg, as well as light elements, for example, Li. Here, we discuss the pros and cons of implementing the ELIISE and ERIN algorithms for resolving unknown ion positions in crystal structures, where traditional structural techniques fail.

ELIISE uses both the electronic charge density (CD) and the local electrostatic potential $LP(r)$ derived from first-principles calculations of the empty organic framework. In combination, the CD and the $LP(r)$ are used to find candidate crystallographic sites for light elements in these materials. However, since the CD and the $LP(r)$ are intrinsically linked (as the latter is derived from atomic positions and the electronic charge density distribution), the independent use of the CD and the $LP(r)$ often arrives at similar predictions for candidate ion sites. Since ELIISE directly identifies coordinating environments—regions surrounded by electronegative atoms of the organic frameworks, ELIISE appears most effective for systems where the organic framework structure of the discharged (ion-inserted) phase is known. This signifies that prior experiments determining the space group, lattice constants, and atom positions of the organic framework (excluding the mobile ions) must be available. In some sense, this limits the capabilities of the ELIISE strategy. For example, if only the charged-phase structure of an organic system is available and significant structural changes and/or phase transitions are expected upon electrochemical reactions, using ELIISE may yield unreliable results.

To address the apparent limitation of ELIISE, we have developed an alternative method called ERIN. ERIN generates a sparse set of points that cover all possible Wyckoff sites, ranked from the electrostatically lowest-energy sites to the highest-energy sites. Therefore, ERIN should be applied when the structure of the organic framework, whose ion-inserted/discharged phases remain unknown. In such cases, the charged (empty) structure of the materials can be used as a starting point, assuming the unit cell of the pristine organic crystal shape and most of the symmetry remains unchanged or structural modifications occur gradually (for example, a displacive phase transition) upon reaction with Li (or sodium).

ELIISE identifies the stationary sites, i.e., the local minima, by populating the energy landscape; these sites are expected to be stationary points for ions in the potential created by the anion framework. In contrast, in ERIN, the identified sites are not necessarily located at energy minima –non-stationary sites, such as saddle points, on the potential energy surface, and may be mobile during structural relaxation. These non-stationary sites can perturb the organic framework during their relaxation and promote structural changes that lead the organic framework toward a more stable discharged structure.

Even if the discharge phase does not involve phase transitions, the specific coordination environments and lattice parameters required to host mobile metal atoms in the discharged organic framework may not be available in the charged organic framework. ERIN systematic electrostatic screening helps identify plausible insertion sites that would be hard to locate with ELIISE alone. Note, however, that ERIN predicts more candidate sites than ELIISE, resulting in a greater number of possible ion-vacancy arrangements to evaluate in the SIIE workflow.

Our study has focused on identifying plausible sites for cation insertion; however, with suitable modifications, these algorithms can also be used to locate anion positions in a crystal structure. For example, in the case of the ELIISE method, local minima in the $ALP(r)$ should be used instead of local maxima. The approach remains the same when using CD. In ERIN, instead of selecting the maximum in the $\mathrm{ALP}(\mathrm{r})$ as a candidate site, the minimum in the $\mathrm{ALP}(\mathrm{r})$ should be chosen.

At its core, ERIN uses the Coulomb electrostatic potential to model the energy landscape and identify sites generated from electronic charge densities obtained from first-principles calculations. Other methods, such as machine learning interatomic potentials [57,67,68] or Bond Valence Sums[69] appear viable for modeling and developing the potential energy landscape for ion insertion.

In general, when testing the ELIISE and ERIN algorithms on 16 previously known structures (**Figure 4** and **Figure S1**), we demonstrated how these algorithms, in combination with the SIIE workflow, offer a reliable and systematic approach for accurately predicting unknown ion positions in organic frameworks.

Another important aspect is the integration of machine learning interatomic potentials (MLIPs), such as MACE, into the SIIE workflow. In principle, ERIN and ELIISE are applicable together with any foundational MLIP model, including ME3GNET, CHGNET, and ORB-3, among others.[57,67,68] This hybrid MLIP–DFT approach appears to be viable, significantly reducing the computational burden of exploring vast combinatorial site occupancies while providing sufficient accuracy in predicting ion positions. Furthermore, the ability of MLIPs to screen thousands of candidate configurations before subsequent, more accurate but computationally intensive refinements using first-principles calculations highlights a scalable pathway for high-throughput exploration of ion-insertion chemistries in soft systems, such as organic crystals, porous systems, metal-organic frameworks, and larger molecules with biological relevance.

In this vein, we have tested MACE predictions against those obtained by DFT on the same systems of **Figure 4**. We have observed that when the organic framework is kept fixed at the known discharged structure, identifying the structure with the lowest energy and correct ion arrangements using (MADE) prediction error within a tolerable range of 0.4 does not require DFT; foundational MLIPs without re-training appear suited for this task. Here, we have only tested MACE.[55,56]

A limitation exists within the SIIE workflow when the number of possible ion combinations generated is in the order of $10^4$ to $10^6$, as this can still require significant computational resources to evaluate total energies across all configurations, even with MLIP such as MACE. In this context, we had relied on the fact that candidate sites are ranked in decreasing order of electrostatic favorability, as indicated by the averaged local potential. Unfavorable sites are systematically excluded during the generation of ion combinations; this removal process continues until the number of valid ion combinations falls below the preset threshold (~$10^3$).

Aside from ELIISE and ERIN, the simultaneous insertion of ions, implemented in the SIIE workflow, is also essential for accurately finding discharged structures, as illustrated by the case study of $Li_4NDC$. In $Li_4NDC$, we proposed Model 3, which yields the lowest DFT total energy (**Figure 3**), and is lower than that of the previously proposed Models 1 and 2.[59] Before this work,[51] other strategies, such as sequential insertion methods, have been proposed.[48,54] In the sequential approach, a new model is created by inserting an atom in a new position in the unit cell, and the structure positions are optimized. This procedure is performed multiple times until the expected stoichiometry is achieved, in the hope of finding a representative global minimum. The sequential approach can lead to incorrect selection of stable candidates because ions may become trapped in local minima of the potential energy surface.

In the case of $Li_4NDC$, determining the optimal ion positions was uncomplicated because the organic structure remained essentially unchanged during lithiation.

However, this is not the case for most systems. For example, when Li$_2$BDC is lithiated, forming Li$_4$BDC, a symmetry reduction is observed.[33] It involves doubling the unit cell along the stacking direction of the organic layers, accompanied by changes in the organic framework.[33] Consequently, attempting to predict ion positions in Li$_4$BDC by starting directly from the Li$_2$BDC structure would produce an inaccurate model.

These cases highlight a broader challenge: it is often unclear whether a significant phase transition occurs during the lithiation process. When this uncertainty arises, a practical approach is first to predict ion positions based on the charged structure (i.e., the host in its oxidized or delithiated form). The resulting model can then be compared with experimental data such as XRD or ND; if the simulated patterns show similarity to the experimental patterns, the predicted structure can serve as a solid starting point for refinement.

However, if the comparison shows significant differences, it becomes necessary to explore alternative structural hypotheses. In such cases, one must generate and assess a set of candidate organic framework motifs to serve as initial models in the ERIN+SIIE workflow. Developing systematic methods for constructing and testing these organic framework variations appears crucial for reliably understanding the structural complexity of ion insertion in molecular crystals.

Besides XRD and ND for determining the structures of organic materials, solid-state nuclear magnetic resonance (ss-NMR) crystallography is another valuable tool that offers structural insights.[70,71] The nascent application of ss-NMR in conjunction with first-principles calculations, termed as NMR crystallography, has been used to establish the structure of functional materials.[72]

## Conclusions

In summary, we have developed a computational framework for accurately predicting ion positions in organic materials, utilizing DFT-derived electrostatic and charge density fields with the ELIISE and ERIN algorithms and the SIIE workflow.

Through case studies of $Na_4C_6O_6$, $Na_2BQ$, and $Li_4NDC$, we showed that ELIISE and ERIN methodologies can reproduce ion positions known from the literature with sub-angstrom accuracy. Additionally, we demonstrate that these methodologies can uncover new, energetically favorable structural models that can improve our understanding of experimental observations.

We have demonstrated that computationally intensive DFT predictions are not a strong requirement for accurately identifying ion positions, and we have shown that foundational machine-learned potentials provide sufficient accuracy for large-scale screening tasks.

Our results demonstrate that ELIISE and ERIN, combined with SIIE, are powerful and can significantly improve, if not augment, the limits of traditional structural characterization techniques. The framework proposed in this paper offers a pathway toward deeper mechanistic insights and the rational design of organic materials for next-generation functional materials. Moreover, MLIPs can efficiently screen thousands of candidate configurations initially, before performing more precise but computationally expensive refinements with first-principles methods. This approach offers a scalable pathway for high-throughput exploration of ion-insertion chemistries in soft systems, including organic crystals, porous materials, metal-organic frameworks, and larger biologically relevant molecules.

## Computational Details

All first-principles calculations presented in this work were performed using the density functional theory (DFT) formalism, as implemented in the Vienna Ab initio Simulation Package (VASP).[73–75] The PAW potentials describe the core electrons. Perdew, Burke, and Ernzerhof (PBE)[76] was used to approximate the unknown DFT exchange and correlation XC functional. To account for Van der Waals interactions, the empirical D3 method proposed by Grimme and collaborators, with Becke-Johnson (BJ) damping, was used.[77,78]

VASP inputs prepared for geometry optimization and energy calculations closely followed the MITRelaxSet[79] as available in pymatgen.[61] The kinetic energy cutoff for the plane waves was set to 520 eV, and the total energy was converged to $10^{-5}$ eV per cell. Geometries (coordinates, volumes, and cell shapes) were considered converged when the forces on all atoms are lower than 0.05 eV/Å. A Γ-centered Monkhorst–Pack[80] grid with a density of 25 $k$-points per $\text{Å}^{-1}$ for all systems. With these DFT settings, local potentials are computed as the sum of Ewald and Hartree parts, excluding the Exchange and Correlation (XC) contributions.

To reduce the number of structures for DFT optimization, a MACE-MP-0 machine-learned foundational model, based on MACE,[55,56] with PFP-based PyTorch implementation of DFT-D3,[77,78,81] was used to evaluate static total energy and/or optimize ion positions. Structural relaxations are carried out until the forces are converged below 0.05 eV/Å.

When MACE is implemented in the SIIE workflow is used as follows:

1. Generate all possible ion combinations. If the number of combinations exceeds a preset limit (~$10^3$), they can be reduced systematically by removing electrostatically unfavorable sites in ELIISE or ERIN before generating possible ion combinations.
2. Use MACE to evaluate all the generated combinations.
3. Select approximately ~$10^2$ of the lowest-energy candidate structures for optimization of ion positions using MACE.
4. From these structures, we choose approximately 10 unique lowest-energy structures for further accurate DFT optimization of the atomic positions.

This hybrid approach, which combines the computational speed of MACE-based MLIPs with the accuracy of DFT, enables us to successfully identify the correct atomic configurations in organic frameworks with substantial structural complexity.

## Data Availability

All the computational data associated with this study, including the input and output files of the simulations, are available on GitHub at https://github.com/caneparesearch/data_site_finder

## Author contributions

**H.R.G.**: Conceptualization, Data curation, Formal analysis, Investigation, Methodology, Validation, Visualization, Writing – original draft, Writing – review & editing. **A.E.L.:** Data curation, Discussion, Writing – original draft, Writing – review & editing. **A.A.P.:** Discussion, Writing – review & editing. **Y.C.:** Discussion, Writing – review & editing. **V. S.C.K.:** Discussion, Writing – review & editing. **L.J.:** Discussion, Writing – review & editing. **J.W.:** Discussion, Writing – review & editing. **Y.C.:**

Discussion, Writing – review & editing. **K.W.:** Discussion, Writing – review & editing. **M.K.Y.C.:** Discussion, Writing – review & editing. **Y.Y.:** Funding acquisition, Supervision, Discussion, Writing – review & editing. **P.C.:** Conceptualization, Funding acquisition, Investigation, Project administration, Resources, Supervision, Visualization, Writing – original draft, Writing – review & editing.

## Acknowledgement

This work is funded by the Energy Storage Research Alliance (ESRA) (DE-AC02-06CH11357), an Energy Innovation Hub funded by the U.S. Department of Energy, Office of Science, Basic Energy Sciences. We are grateful for the support of the Research Computing Data Core at the University of Houston for assistance with the calculations carried out in this work. This research utilized resources of the National Energy Research Scientific Computing Center (NERSC), a Department of Energy User Facility, supported by the NERSC award BES-ERCAP0031978. P.C. acknowledges the Welch Foundation under grants L-E-001-19921203, and E-2227-20250403. Work performed at the Center for Nanoscale Materials, a U.S. Department of Energy Office of Science User Facility, was supported by the U.S. DOE, Office of Basic Energy Sciences, under Contract No. DE-AC02-06CH11357.